%% file: gaussianPQS.tex
\documentclass[aps,pra,twocolumn,groupedaddress]{revtex4-1}
\pdfoutput=1
\usepackage[T1]{fontenc}
\usepackage[utf8]{inputenc}
\usepackage[english]{babel}
\usepackage{bm}	%For bold math symbols
\usepackage{amsopn}	%For DeclareMathOperator command
\usepackage{amssymb}	%For hollow letters (real and complex numbers)
\usepackage{amsmath}	%For intext \smallmatrix
\usepackage{dsfont}		%For identity operator
\usepackage{braket}		%For easy bra and ket
\usepackage{physics}
\usepackage{graphicx}
\usepackage{widebar}
\usepackage{svg}
\usepackage{booktabs}	%For tabular
\usepackage{makecell}	%For multi-line table cells
\usepackage{tabularx}	    %For tabularx
\usepackage{hyperref}
\usepackage{natbib}

\usepackage{tikz}
\usetikzlibrary{arrows,positioning}
\usepackage{environ}
\usepackage{varwidth}

\newcommand{\pr}{\operatorname{Pr}}

\renewcommand{\tr}[1]{\Tr(#1)}

\newcommand{\ang}[1]{\langle #1\rangle}	%angled brackets
\newcommand{\transp}{^{\intercal}}	%transpose symbol

\newcommand{\idop}{\mathds{1}}
\newcommand{\ct}{\widetilde{C}}	%Definition of Ctilde operator

\newcommand{\mean}[1]{\overline{#1}}

\begin{document}

%\preprint{}

\title{Prediction and retrodiction with continuously monitored Gaussian states}

\author{Jinglei Zhang}
\author{Klaus M{\o}lmer}

\affiliation{Department of Physics and Astronomy, Aarhus University, Ny Munkegade 120, DK-8000 Aarhus C, Denmark}

\date{\today}

\begin{abstract}
Gaussian states of quantum oscillators are fully characterized by the mean values and the covariance matrix of their quadrature observables. We consider the dynamics of a system of oscillators subject to interactions, damping, and continuous probing which maintain their Gaussian state property. Such dynamics is found in many physical systems that can therefore be efficiently described by the ensuing effective representation of the density matrix $\rho(t)$. Our probabilistic knowledge about the outcome of measurements on a quantum system at time $t$ is not only governed by $\rho(t)$ conditioned on the evolution and measurement outcomes obtained until time $t$, but is also modified by any information acquired after $t$. It was shown in [Phys. Rev. Lett. {\bf 111}, 160401 (2013)] that this information is represented by a supplementary matrix, $E(t)$. We show here that the restriction of the dynamics of $\rho(t)$ to Gaussian states implies that the matrix $E(t)$ is also fully characterized by a vector of mean values and a covariance matrix. We derive the dynamical equations for these quantities and we illustrate their use in the retrodiction of measurements on Gaussian systems.
\end{abstract}

\maketitle

\section{Introduction}

We describe the state of a quantum system by a wave function $\psi$ or, more generally for an open system, by a density matrix $\rho$. If the system is subject to repeated or continuous measurements, $\rho(t)$ is evolved by a combination of unitary and dissipative dynamics, and it is subject to measurement back action that is dependent on the random outcome of the quantum measurements performed. While we commonly refer to $\rho(t)$ as the {\em state} of the system, we do not as physicists generally agree on its precise physical meaning. We do agree, however, that $\rho(t)$ provides, and is also fully specified by, the probabilities for the outcomes of all possible measurements on the system at the time $t$.

In our daily lives, we often encounter situations where we acquire information that refines or fundamentally changes our knowledge about past events and what we earlier held to be true. Similarly, in a quantum physics experiment, we may ask ourselves what are the possible and most likely outcomes of a past measurement, conditioned on both our earlier and later observation of the system. A pair of matrices $\rho(t)$ and $E(t)$ were shown in \cite{gammelmark_past_2013} to yield the retrodicted probabilities for any measurement at a past time $t$, and was hence labelled the {\em past quantum state}, in analogy to the usual quantum state $\rho(t)$ which provides the probabilities for any measurement at the present time $t$. The past quantum state formalism has been used to retrodict the photon number distribution of a microwave cavity field subject to probing by transmission of atoms \cite{rybarczyk_forward-backward_2015}, and its exploitation of the full measurement record to retrodict the outcomes of past measurements has been applied and tested in a series of experiments on superconducting qubits \cite{campagne-ibarcq_observing_2014,weber_mapping_2014, tan_prediction_2015, tan_quantum_2016}. Better knowledge of the time evolution of a quantum system may yield better estimates of unknown parameters \cite{gammelmark_hidden_2014,tsang_optimal_2010, tsang_fundamental_2012}, and it may offer insight into the  temporal correlations in measurement records \cite{wiseman_weak_2002,murch_observing_2013,xu_correlation_2015,guevara_quantum_2015,foroozani_correlations_2016}.

The past quantum state theory is derived from the POVM formalism and the quantum theory of measurements. It has been so far implemented in the form of a stochastic master equation for the density operator $\rho$ and a corresponding adjoint equation evolving the effect matrix $E$ backwards in time \cite{gammelmark_past_2013}. However, for many continuous variable systems such as multi-mode light fields \cite{gardiner_quantum_2004, braunstein_quantum_2005} optical and optomechanical systems \cite{genoni_quantum_2015, wieczorek_optimal_2015}, cold  atomic ensembles \cite{molmer_estimation_2004}, and Bose-Einstein condensates \cite{wade_squeezing_2015}, the most general density operator treatment is inconvenient and may not be needed as the system dynamics may be restricted to Gaussian states, i.e., states for which the Wigner function is a Gaussian function of the quadrature coordinates. Under conditions that are often fulfilled in experiments, the Gaussianity of such systems is preserved and, although they have infinite dimensional Hilbert spaces, their dynamical evolution is fully described by the evolution of the mean values and the covariances of the quadrature observables.

In this article we develop a past quantum state theory for Gaussian states in which both $\rho$ and $E$ allow an effective representation by mean values and covariances of the quadrature observables. We derive and show how to solve the equations of motion for the vectors of mean values and covariance matrices, and we demonstrate their use in the retrodiction of past probability distributions for measurements of quadrature observables. 

In Sec. \ref{sec:pqs} we recall the general past quantum state formalism. In Sec. \ref{sec:markovian} we specialize to the case of Gaussian states and measurements that preserve their Gaussian character. In Sec. \ref{sec:applications} we discuss and illustrate the formalism and its applications to simple examples. Sec. \ref{sec:conclusion} concludes the work.

\section{Past quantum state}
\label{sec:pqs}

Consider a quantum system that is described at time $t$ by the density operator $\rho(t)$.  A generalized measurement is described by a POVM, i.e., by a set of operators $\{\hat{M}_m\}$ which obeys the normalization $\sum_m \hat{M}_m^\dagger \hat{M}_m = \idop$, and gives the probability of measuring outcome $m$ by the generalized Born rule:
\begin{equation}
\label{eq:born}
\pr(m,t) = \tr{\hat{M}_m \rho(t) \hat{M}_m^\dagger}.
\end{equation}
After a measurement has been performed, the state is conditioned on the measurement outcome and is updated according to  $\rho(t) \rightarrow \hat{M}_m \rho(t) \hat{M}_m^\dagger/\pr(m,t)$.

If after time $t$ the system is further probed until some final time $T>t$, more information is accumulated about the system, and the probability of outcome $m$ at time $t$ conditioned on the later measurements may generally differ from Eq. (\ref{eq:born}). As shown in  \cite{gammelmark_past_2013}, the past probability for the outcome $m$ conditioned on previous and later probing of the system can be written as
\begin{equation}
\label{eq:pp}
\pr_\text{p}(m,t) = \frac{\tr{\hat{M}_m \rho(t) \hat{M}_m^\dagger E(t)}}{\sum_{m'}\tr{\hat{M}_{m'} \rho(t) \hat{M}_{m'}^\dagger E(t)}}.
\end{equation}
Here, the density matrix $\rho(t)$ represents the prior information about the system while the so-called effect matrix $E(t)$ serves as a quantum generalization of Bayes' rule and updates the outcome probabilities in Eq. (\ref{eq:born}) by the data retrieved between $t$ and $T$. The density matrix $\rho(t)$ is generally found by solving a stochastic master equation since it depends on the random measurement outcomes obtained before time $t$. Similarly, $E(t)$ is found by solving an adjoint master equation backward in time from the final time condition $E(T)=\idop$ and it depends on the measurements performed between time $t$ and $T$. If no measurement is performed on the system after time $t$, the effect operator $E(t)$ equals the identity operator, and  Eq. (\ref{eq:pp}) coincides with the usual Born rule in Eq. (\ref{eq:born}). Assuming $T>t$, we refer to the pair of matrices $(\rho(t),E(t))$, conditioned on the full measurement record from the time interval $[0,T]$, as the past quantum state.

\section{Markovian evolution for a Gaussian state}
\label{sec:markovian}

Harmonic oscillators are continuous variable systems with infinite dimensional Hilbert spaces and density matrices, and characterization of their quantum state by mean values and covariances hence represents a major simplification. This simplification has been successfully applied both to Gaussian states and operations where it is exact and to physical systems where quantum fluctuations are well approximated by Gaussian distributions. 

\begin{figure}
\centering
\def\svgwidth{\columnwidth}\footnotesize
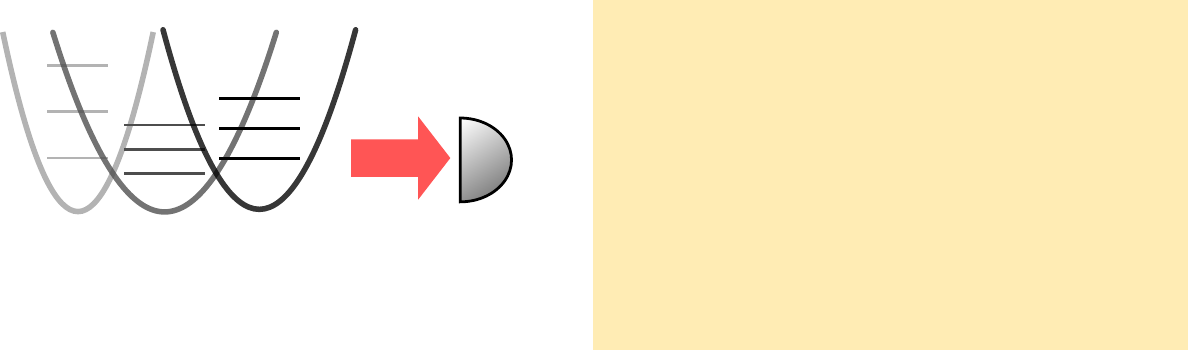
\caption{An open system of oscillators is subject to continuous weak probing or monitoring of its leakage of excitation into surrounding bath degrees of freedom. The measurements yield a stochastic measurement record. The evolution of the density matrix $\rho(t)$ and the effect matrix $E(t)$ is conditioned on the measurement signal acquired in the time intervals $[0,t)$ and $(t,T]$, respectively.}
\label{fig:system}
\end{figure}

We consider the density matrix evolution of a Gaussian state subject to a Hamiltonian and to a Markovian coupling to reservoir oscillators that are quadratic in the system and reservoir quadrature observables. We include also the decoherence and measurement back action imposed by continuous probing of either the system or reservoir quadrature degrees of freedom, see Fig. \ref{fig:system}. This evolution will maintain a Gaussian state of the oscillators and permits an effective description of the system evolution in terms of stochastically evolving mean values and a deterministically evolving  covariance matrix. In this section we recall the derivation of this description for the state $\rho(t)$ and we derive the similar representation and formalism to be applied to the effect matrix $E(t)$.

%\begin{figure}
%{\centering
%\includesvg[width=\columnwidth,svgpath=images/,pretex=\footnotesize]{system}}
%\caption{An open system of oscillators is subject to continuous weak probing or monitoring of its leakage of excitation into surrounding bath degrees of freedom. The measurements yield a stochastic measurement record. The evolution of the density matrix $\rho(t)$ and the effect matrix $E(t)$ is conditioned on the measurement signal acquired in the time intervals $[0,t)$ and $(t,T]$, respectively.}
%\label{fig:system}
%\end{figure}

\subsection{Forward evolution of the quantum state}

Let us consider a system consisting of $n$ oscillator modes described by the quadrature operators $\vb{r}\transp = (q_1,p_1,\dots,q_n,p_n)$ that obey the canonical commutation relation $\comm{q_j}{p_k} = i\delta_{jk}$ $(\hbar = 1)$, and let us for notational convenience define the symplectic matrix $\Omega$ with elements $i\Omega_{jk}=\comm{r_j}{r_k}$. We assume that the Hamiltonian is quadratic in the quadrature operators $H=\frac{1}{2}\sum_{jk} R_{jk} r_j r_k$, where $R$ is a real symmetric matrix. Furthermore, the oscillators are in contact with a Markovian bath and they may be subject to weak probing of observables $\{c_h\}$ ($h=1,\dots,m$) that are linear in the system quadrature operators.
The average behavior of the system is then described by an unconditioned density matrix, which solves the master equation
\begin{equation}
\label{eq:averagedrho}
\dd{\rho} = -i\comm{H}{\rho}\dd{t} + \sum_h \mathcal{D}[c_h]\rho \dd{t},
\end{equation}
where the dissipation superoperator $\mathcal{D}[c]\rho = c\rho c^\dagger - \frac{1}{2}\acomm{c^\dagger c}{\rho}$ is of the so-called Lindblad form. If $c=\sqrt{\gamma}a$, where $a$ is an annihilation operator for an oscillator, $\mathcal{D}[c]\rho$ describes damping with rate $\gamma$ of that oscillator, while if $c=\sqrt{\kappa}(a+a^\dagger)$, it describes the decoherence due to the dispersive interaction of the oscillator quadrature $q=a+a^\dagger$ with a quantum probe of amplitude $\propto \sqrt{\kappa}$. If a dissipation channel is monitored through either homodyne or heterodyne detection, a measurement current is obtained and it can be written as $\dd{Y}(t) = \sqrt{\eta} \tr{\rho({c} + {c}^\dagger)} \dd{t} + \dd{W}(t)$, where $0\leq\eta\leq 1$ denotes the measurement efficiency and $W(t)$ is a stochastic Wiener process. The back action due to the measurement is represented by adding a stochastic term $\sqrt{\eta}(c \rho + \rho c^\dagger) \dd{Y}(t)$ to the master equation for $\rho(t)$ given in Eq. (\ref{eq:averagedrho}). The detector efficiency $\eta$ assumes the value zero if a channel is left unmonitored. We can thus represent dissipation and measurements on an equal footing, and we obtain a stochastic master equation for the density matrix \cite{wiseman_quantum_2010}
\begin{equation}
\begin{split}
\label{eq:ulindblad}
\dd{\rho} = -i\comm{H}{\rho}\dd{t} &+ \sum_h \mathcal{D}[c_h]\rho \dd{t} +\\
&+ \sum_h \sqrt{\eta_h}(c_h \rho + \rho c_h^\dagger) \dd{Y_h}(t).
\end{split}
\end{equation}

A trace-preserving form of Eq. \ref{eq:ulindblad} can be written as \cite{wiseman_quantum_2010}
\begin{equation}
\begin{split}
\label{eq:lindblad}
\dd{{\rho}} = -i\comm{H}{\rho}\dd{t} &+ \sum_h \mathcal{D}[c_h]\rho \dd{t} + \\
&+\sum_h \sqrt{\eta_h} \mathcal{H}[c_h]\rho \dd{W_h}(t),
\end{split}
\end{equation}
where the measurement superoperator $\mathcal{H}$ is defined as $\mathcal{H}[c]\rho = c\rho + \rho c^\dagger -\Tr(\rho(c+ c^\dagger))$. We note that homodyne detection of a quadrature of the radiation emitted from an oscillator, $c =\sqrt{\gamma}a$, has a different back action than the probing of the hermitian oscillator quadrature $a+a^\dagger$. Both, however, preserve Gaussian states.

For any operator, and in particular for the density operator $\rho$, it is possible to associate a Wigner function \cite{hillery_distribution_1984} $W_\rho({\bf r})$ on the multi-dimensional position-momentum phase space. The Wigner function is a quasiprobability distribution that provides a description of the system equivalent to the density operator formalism. One can show \cite{gardiner_quantum_2004} that the action of the quadrature operators on the density matrix can be represented by multiplication and first order derivatives with respect to the phase space coordinates, namely
\begin{equation}
\label{eq:phaseevolution}
\medmuskip=0mu
\thinmuskip=0mu
\thickmuskip=0mu
\begin{split}
&q_j\rho \rightarrow \left(q_j + \frac{i}{2}\partial_{p_j}\right) W_\rho({\bf r}), \qquad  p_j \rho \rightarrow \left(p_j - \frac{i}{2}\partial_{q_j}\right) W_\rho({\bf r }), \\
&\rho q_j \rightarrow \left(q_j - \frac{i}{2}\partial_{p_j}\right)W_\rho({\bf r}), \qquad \rho p_j \rightarrow \left(p_j + \frac{i}{2}\partial_{q_j}\right)W_\rho({\bf{r}}).
\end{split}
\end{equation}
The stochastic master equation for $\rho$ can thus be transformed into a differential equation for $W_\rho({\bf r})$.

\begin{figure}
\centering
\resizebox{\columnwidth}{!}{
\input{plan.tex}}
\caption{The phase space formalism allows a Gaussian density operator to be simply described by the first and second moments of its Wigner function. In general the conditional evolution of a density operator $\rho$ suggests a corresponding backward evolution for the so-called effect matrix $E$.  The Wigner function associated with the effect matrix $E$ is also fully characterized by first and second moments solving similar equations as the moments for $\rho$.}
\label{fig:scheme}
\end{figure}
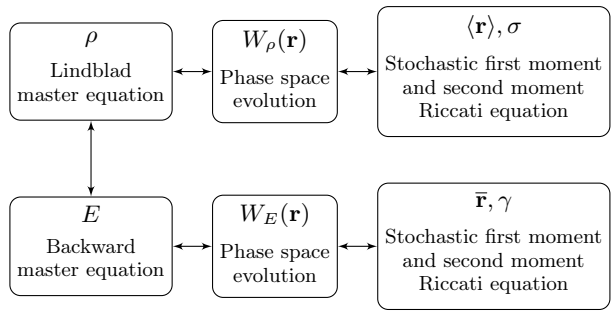

If the Hamiltonian $H$ is at most quadratic and if the operators $\{c_h\}$ representing dissipation and monitoring are linear in the quadrature observables, it is easy to see that the stochastic master equation leads to a Fokker-Planck  equation of evolution for the Wigner function, involving first and second order derivatives that maintain the Gaussian form, i.e., the state of the system is at all times represented by a Gaussian phase space function. Therefore the dynamics of the density matrix is completely described by the evolution of the first and second statistical moments of the quadrature coordinates. For any finite number of oscillator modes, this yields a vast reduction compared to the infinite dimension of the system Hilbert space (see Fig. \ref{fig:scheme}).

We define the vector of first moments, $\ang{\vb{r}} \equiv \tr{\vb{r}\rho}$, and the covariance matrix ${\sigma}$, $\sigma_{jk} \equiv \ang{\acomm{{r}_j}{{r}_k}} - 2\ang{{r}_j}\ang{{r}_k}$. The evolution of the system is now completely characterized by the following equations, derived in the Appendix \ref{sec:derivation}:

\begin{equation}
\label{eq:firstmom}
\dd{\ang{\vb{r}}}  = A \ang{\vb{r}} \dd{t} + (\sigma B\transp - N\transp) \sqrt{\bm{\eta}}\dd{{\vb W}(t)},
\end{equation}
\begin{equation}
\label{eq:riccati}
\begin{split}
\dv{\sigma}{t} %&\equiv \frac{{\sigma}(t+\dd{t}) - {\sigma}(t)}{\dd{t}}= \\
&= A \sigma + \sigma A\transp + D - 2(\sigma B\transp - N\transp) \bm{\eta} (\sigma B\transp - N\transp)\transp,
\end{split}
\end{equation}
where we have defined the vector of Wiener increments $\dd{\vb{W}}\transp \equiv (\dd{W_1},\dots,\dd{W_m})$ and the diagonal matrix of efficiencies $\bm{\eta} \equiv \operatorname{diag}(\eta_1,\dots,\eta_m)$. The drift and diffusion matrices $A$ and $D$, defined in the Appendix \ref{sec:derivation}, describe the average evolution of the system independent of the measurement record, while the matrices $B$ and $N$ represent the back action due to the measurements.
It is worth noting that while the evolution of the first moments is stochastic and depends on the outcomes of the measurements through the Wiener noise $\dd{{\vb W}}$, the evolution of the covariance matrix of Gaussian states is deterministic and is described by a non-linear Riccati matrix equation.

\subsection{Backward evolution of the effect matrix}
As shown in \cite{gammelmark_past_2013}, the effect matrix $E(t)$ is evolved backwards in time by the Hilbert-Schmidt adjoint of master equation (\ref{eq:ulindblad}) for the density operator, namely
\begin{equation}
\label{eq:bulindblad}
\begin{split}
\dd{E} &\equiv \dd{E}(t-\dd{t}) - \dd{E}(t) = \\
&= i\comm{H}{E}\dd{t} + \sum_h \mathcal{D}^\dagger[c_h]E \dd{t} +\\
&\hphantom{= i\comm{H}{E}\dd{t} }+ \sum_h \sqrt{\eta_h}(c_h^\dagger E + E c_h) \dd{Y_h}(t-\dd{t}),
\end{split}
\end{equation}
where, if $\mathcal{C}X = LXR$, the adjoint superoperator is $\mathcal{C}^\dagger X = L^\dagger X R^\dagger$.
Eq. (\ref{eq:bulindblad}) does not preserve the trace of the effect matrix. For direct numerical applications this does not constitute a problem, since Eq. (\ref{eq:pp}) for the retrodicted probabilities explicitly includes a renormalization factor. In the present work, however, we are interested in applying a phase space representation of the effect matrix and its first and second moments can only be obtained from a normalized phase space distribution. We therefore convert Eq. (\ref{eq:bulindblad}) to the following, trace-preserving master equation:
\begin{equation}
\label{eq:blindblad}
\begin{split}
\dd{E} = i\comm{H}{E}\dd{t} +&\sum_{h}\left(\mathcal{D}^\dagger[c_h]E - \left( \mean{{c}_h {c}_h^\dagger} - \mean{{c}_h^\dagger {c}_h}\right)E\right)\dd{t} +\\
+&\sum_{h}\sqrt{\eta_h}\mathcal{H}[c_h^\dagger]E \dd{s}_h(t-\dd{t}),\\
\end{split}
\end{equation}
where we have defined the notation $\mean{c} \equiv \tr{c E}$. The action of the adjoint of the dissipation superoperator $\mathcal{D}$ is specified as $\mathcal{D}^\dagger[c]E = c^\dagger E c - \frac{1}{2}\acomm{c^\dagger c}{E}$, while the measurement superoperator appears in the same form as in the evolution of $\rho$, but with the adjoint argument, $\mathcal{H}[c^\dagger]E = c^\dagger E + E c -\Tr(E(c^\dagger +c))$. The stochastic terms, incorporating the outcome $dY_{h}(t)$ of the measurements, are defined as $\dd{s_h}(t) = \dd{Y_h}(t) - \sqrt{\eta_h}( \mean{{c}_h^\dagger + {c}_h} )\dd{t}$, and we group them into the vector $\dd{\vb{s}}\transp \equiv (\dd{s_1},\dots,\dd{s_m})$.  Note that $\rho(t)$ can be determined without any knowledge of $E(t)$ and measurement outcomes after $t$. Similarly, $E(t)$ can be computed without knowledge of $\rho(t)$ and any measurement data before $t$.

In order to represent and study the evolution of $E$ on the position-momentum phase space it is sufficient to recall that, as for the Wigner function of the density matrix $\rho$ (see Eq. (\ref{eq:phaseevolution})), all superoperators acting on $E$ correspond to appropriate first and  second order differential operators acting on the Wigner function of $E$. We conclude that under the same conditions both the evolution of $\rho$ and $E$ preserve the Gaussian character of their phase space Wigner distributions, and they are thus both fully characterized by their first and second moments.

Let $\mean{\vb{r}} = \tr{\vb{r}E}$ and $\gamma_{jk} = \mean{\acomm{r_j}{r_k}} -2 \mean{r_j}\mean{r_k}$ denote the vector of mean values and the covariance matrix of $E$ respectively. As shown in the Appendix \ref{sec:derivation}, we obtain the backward equations of evolution of these quantities:

\begin{equation}
\label{eq:bfirst}
\dd{\mean{\vb{r}}} \equiv {\mean{\vb{r}}}(t-\dd{t}) - {\mean{\vb{r}}}(t) = -A\mean{\vb{r}}\dd{t} + (\gamma B\transp + N\transp) \sqrt{\bm{\eta}}\dd{\vb{s}},
\end{equation}
\begin{equation}
\label{eq:briccati}
\begin{split}
\dv{\gamma}{t} &\equiv \frac{{\gamma}(t-\dd{t}) - {\gamma}(\dd{t})}{\dd{t}} =\\
&= -A\gamma - \gamma A\transp + D - 2(\gamma B\transp + N\transp) \bm{\eta} (\gamma B\transp + N\transp)\transp.
\end{split}
\end{equation}

Compared to the evolution of $\sigma$ in Eq. (\ref{eq:riccati}), $\gamma$ evolves with the opposite drift matrix $A$ and the same diffusion matrix $D$. The case where all the detection mode operators $c_h$ are Hermitian is worth mentioning: in this case $N=0$, see Eq. (\ref{eq:rhofirst}) in the Appendix \ref{sec:derivation}, and therefore the measurement contribution to the evolution of $\gamma$ given in Eq. (\ref{eq:briccati}) becomes the same as for $\sigma$ in Eq. (\ref{eq:riccati}). This also follows from the original stochastic master equations, where the superoperator $\mathcal{H}[c]$ acting on  $\rho$ coincides with $\mathcal{H}[c^\dagger]$ acting on $E$, when $c=c^\dagger$.

Despite the symmetry and similarity between the moments $(\ang{r}, \sigma)$ associated to $\rho$ and $(\mean{r}, \gamma)$ associated to $E$, it is important to remember that the effect matrix has a very different physical meaning compared to the density operator: $E(t)$ represents a Bayesian update of our prior probabilistic knowledge $\rho(t)$, depending on the outcomes of the later measurements. It does not in itself yield a probability of measurement outcomes (unless $\rho$ is the identity matrix). 

\section{Applications}
\label{sec:applications}

We have developed a complete theory of Gaussian past quantum states, giving access to the first and second moments of the phase space distribution for the operators $\rho$ and $E$. We can, in principle, convert the information represented by $(\ang{r},\sigma)$ and $(\mean{r},\gamma)$ to phase space Wigner functions and, subsequently to, e.g., a Fock state matrix representation of the operators. For most applications this will not be necessary, and it is certainly not a practical approach to deal with the predictions by the theory.

If, for example, the measurement we want to retrodict is a projective measurement on orthogonal states, $\hat{M} = \hat{M}^\dagger = \ketbra{m}$, the past probability expression in Eq. (\ref{eq:pp}) yields
\begin{equation}
\label{eq:projectivepp}
\pr_\text{p}(m) \propto \ev{\rho}{m} \ev{E}{m}.
\end{equation}

The important corrections to our predictions due to the effect matrix $E$ may be evaluated by treating the term $\ev{E}{m}$ as if $E$ is the density matrix of a Gaussian state, and by calculating the matrix element in the same way as with a density matrix $\rho$. For that purpose we may use a wealth of results from quantum optics about the outcome of measurements on Gaussian states. The Fock state content of squeezed and displaced states are for example available in the literature (see \cite{dodonov_multidimensional_1994} and references therein).

Working with Gaussian phase space distributions, it is particularly easy to obtain the probability distributions for the measurement of position and momentum degrees of freedom and their linear combinations. Let us consider a single oscillator, and indicate with $\ket{x,\theta}$ the eigenvectors of the quadrature operator ${x}_\theta=q\cos\theta+p\sin\theta$. Then the past probability Eq. (\ref{eq:pp}) of a measurement of ${x}_\theta$ reads
\begin{equation}
\label{eq:pastx}
\pr_\text{p}(x,\theta)=\frac{\ev{\rho}{x,\theta}\ev{E}{x,\theta}}{\int_{-\infty}^{\infty}\dd{x'}\ev{\rho}{x',\theta}\ev{E}{x',\theta}}.
\end{equation}

Due to the Gaussian form of the Wigner functions for $\rho$ and $E$, the marginal distributions  $\ev{\rho}{x,\theta}$ and $\ev{E}{x,\theta}$ are also Gaussian distributions, and so is their product in the numerator of Eq. (\ref{eq:pastx}).

A single oscillator with a Gaussian density operator is characterized by mean values  $\ang{{\vb r}}\transp = (\ang{q},\ang{p})$ and covariance matrix $\sigma = \left(\begin{smallmatrix} \sigma_{qq} & \sigma_{qp} \\ \sigma_{qp} & \sigma_{pp}\end{smallmatrix}\right)$, while its Gaussian effect matrix is characterized by $\mean{r} = (\mean{q},\mean{p})\transp$ and covariance matrix $\gamma = \left(\begin{smallmatrix} \gamma_{qq} & \gamma_{qp} \\ \gamma_{qp} & \gamma_{pp}\end{smallmatrix}\right)$. The past probability for the measurement of a generic quadrature $x_\theta$ is then a Gaussian distribution with average $x_{\theta,\text{p}}$ and variance $\operatorname{Var}(x_{\theta,\text{p}})=\Delta(x_{\theta,\text{p}})/2$ given by
\begin{equation}
\label{eq:polarvariance}
\begin{split}
x_{\theta,\text{p}} &= \frac{(\ang{q}\cos{\theta}  + \ang{p}\sin\theta)\gamma_{\theta} + (\mean{q}\cos\theta  + \mean{p}\sin\theta )\sigma_\theta}{\gamma_\theta + \sigma_\theta},\\
&\frac{1}{\Delta(x_{\theta,\text{p}})} = \frac{1}{\sigma_\theta} + \frac{1}{\gamma_\theta},
\end{split}
\end{equation}
where we have defined $\sigma_\theta=  \sigma_{qq}\cos^2\theta - 2\sigma_{qp}\sin\theta \cos\theta  +  \sigma_{pp}\sin^2\theta$ and analogously for $\gamma_\theta$.

\begin{figure}
\centering
\def\svgwidth{\columnwidth} \footnotesize
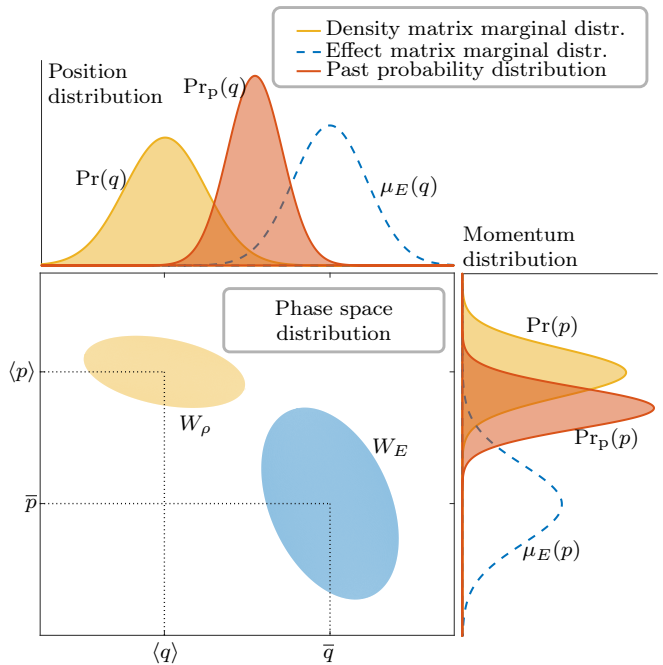
\caption{A Gaussian past quantum state is represented in phase space by the Wigner functions for the density matrix $\rho$ and the effect matrix $E$. Their covariance matrix ellipses project onto Gaussian marginal distributions of the separate quadratures and the marginals of $W_\rho$ give exactly the probability distribution for quadrature measurements. The product of the marginal distributions yields the retrodicted probability distributions, cf. Eq. (\ref{eq:projectivepp}), which are also Gaussian functions with the mean and variance given in Eq. (\ref{eq:pastprediction}).}
\label{fig:phasespace}
\end{figure}

For the position quadrature measurement ($\theta=0$) the past distribution is characterized by the position average $q_\text{p}$ and variance $\operatorname{Var}(q_\text{p})=\Delta(q_\text{p})/2$, given by
\begin{equation}
\label{eq:pastprediction}
q_\text{p} = \frac{\ang{q}\gamma_{qq} + \mean{q}\sigma_{qq}}{\sigma_{qq} + \gamma_{qq}}, \qquad \frac{1}{\Delta(q_\text{p})} = \frac{1}{\sigma_{qq}}+\frac{1}{\gamma_{qq}}.
\end{equation}
The variance formula shows that the incorporation of any information from measurements after time $t$, represented by $\gamma_{qq}$, reduces the uncertainty on the retrodiction of the value of $q$.
Figure \ref{fig:phasespace} shows Gaussian Wigner distributions for $\rho$ and $E$ and their marginal $q$ and $p$ distributions together with the retrodicted marginal distributions.

%\begin{figure}
%\centering
%\includesvg[width=\columnwidth,svgpath=images/,pretex=\footnotesize]{phasespace_new}
%\caption{A Gaussian past quantum state is represented in phase space by the Wigner functions for the density matrix $\rho$ and the effect matrix $E$. Their covariance matrix ellipses project onto Gaussian marginal distributions of the separate quadratures and the marginals of $W_\rho$ give exactly the probability distribution for quadrature measurements. The product of the marginal distributions yields the retrodicted probability distributions, cf. Eq. (\ref{eq:projectivepp}), which are also Gaussian functions with the mean and variance given in Eq. (\ref{eq:pastprediction}).}
%\label{fig:phasespace}
%\end{figure}

\subsection{Decaying oscillator subject to homodyne detection}
As a simple application of the past quantum state analysis to a Gaussian system let us consider a single oscillator mode with Hamiltonian $H=\Omega a^\dagger a$, causing oscillation at frequency $\Omega$ of the oscillator quadratures, and interacting dissipatively with a zero temperature bath leading to decay of the oscillator with rate $\Gamma$. This could represent the case of a Gaussian state of light, leaking out of a cavity, and we imagine that the emitted field is subject to homodyne detection. For the initial state of the system we assume a displaced thermal state.

\begin{figure}
\centering
\def\svgwidth{\columnwidth} \footnotesize
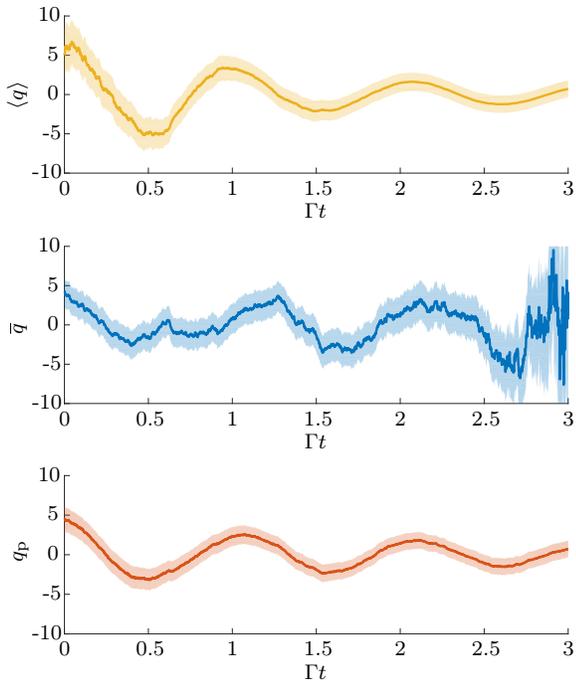
\caption{Time evolution of: (a) forward first moment $\ang{q}$, (b) backward first moment $\overline{q}$, (c) past estimation of position operator $q_\text{p}$. The shaded areas in the two upper panels indicate the values given by $\sqrt{\sigma_{qq}}$ and $\sqrt{\gamma_{qq}}$ respectively, while the corresponding uncertainty on the retrodicted value is given by $(\sigma_{qq}^{-1}+\gamma_{qq}^{-1})^{-\frac{1}{2}}$ (see Eq. (\ref{eq:pastprediction})). The harmonic oscillator has frequency $\Omega$ and damping constant $\Gamma$ where $\Omega/\Gamma=6$, while the efficiency of the homodyne detection is $\eta=0.5$. The initial state is a thermal state with initial first moment $\vb{r}(0)\transp=(5,0)$ and covariance matrix $\sigma(0)=10 \times \idop$.}
\label{fig:oscillator}
\end{figure}

The dynamics of the system can be solved using Eq. (\ref{eq:firstmom}) and (\ref{eq:riccati}) to obtain a full characterization of $\rho(t)$ and Eq. (\ref{eq:bfirst}) and (\ref{eq:briccati}) for $E(t)$. With the resulting $(\ang{q},\sigma_{qq})$ and $(\mean{q},\gamma_{qq})$ one can then calculate the past estimation of the position quadrature of the oscillator at any time in the interval $[0,T]$ by equation (\ref{eq:pastprediction}). The results are shown in Fig. \ref{fig:oscillator}, where we see how the oscillating quadrature $q$ has a mean value that is governed by the unitary evolution of the harmonic oscillator, the dissipation into the environment, and also by the random measurement outcome from the homodyne detection. The variance $\sigma_{qq}$, on the other hand, decreases with time (upper panel). The middle panel shows the similar stochastic and deterministic evolution of $(\mean{q},\gamma_{qq})$ for the effect matrix, while the last panel shows our retrodicted knowledge about the position quadrature of the system. Compared to the usual forward evolution given by $\rho$, by including in the analysis the full past quantum state, the noise on our estimate (represented by the shaded area) is smaller at all times.

%\begin{figure}
%\centering
%\includesvg[width=\columnwidth,svgpath=images/,pretex=\footnotesize]{oscillator}
%\caption{Time evolution of: (a) forward first moment $\ang{q}$, (b) backward first moment $\overline{q}$, (c) past estimation of position operator $q_\text{p}$. The shaded areas in the two upper panels indicate the values given by $\sqrt{\sigma_{qq}}$ and $\sqrt{\gamma_{qq}}$ respectively, while the corresponding uncertainty on the retrodicted value is given by $(\sigma_{qq}^{-1}+\gamma_{qq}^{-1})^{-\frac{1}{2}}$ (see Eq. (\ref{eq:pastprediction})). The harmonic oscillator has frequency $\Omega$ and damping constant $\Gamma$ where $\Omega/\Gamma=6$, while the efficiency of the homodyne detection is $\eta=0.5$. The initial state is a thermal state with initial first moment $\vb{r}(0)\transp=(5,0)$ and covariance matrix $\sigma(0)=10 \times \idop$.}
%\label{fig:oscillator}
%\end{figure}

\subsection{Retrodiction beyond the Heisenberg uncertainty relation}
Heisenberg's uncertainty relation states a fundamental limitation to how well one can predict the outcome of the measurements of two non-commuting observables. States may exist for which one measurement can be very precisely predicted, but then the other observable will be correspondingly less well predicted. The general uncertainty relation applies to any pure or mixed state of quantum systems but it is only concerned with the prediction of future measurements, and it does not describe our ability to retrodict what was the outcome of a measurement on a system at a past time $t$, if we have access to the system both before and after that measurement.

For any two observables $\hat{A}$ and $\hat{B}$, we may thus prepare an eigenstate of $\hat{A}$ before $t$, and we may perform a projective measurement of $\hat{B}$ right after $t$, and we can then with certainty retrodict the outcome of the measurement of any of the two observables, even when $\hat{A}$ and $\hat{B}$ do not commute: Clearly, if $\hat{A}$ was measured, the result would have to match the initially prepared state, while if $\hat{B}$ was measured, the result must have been the same as what we obtained with our subsequent measurement of the same operator. Vaidman et al. \cite{vaidman_how_1987} for example, proposed to prepare a spin in a $\sigma_x$ eigenstate and subsequently detect $\sigma_z$. Therefore we can with certainty retrodict the outcome of a measurement of both $\sigma_x$ and $\sigma_z$ performed at an intermediate time.

\begin{figure}
\centering
\def\svgwidth{\columnwidth} \footnotesize
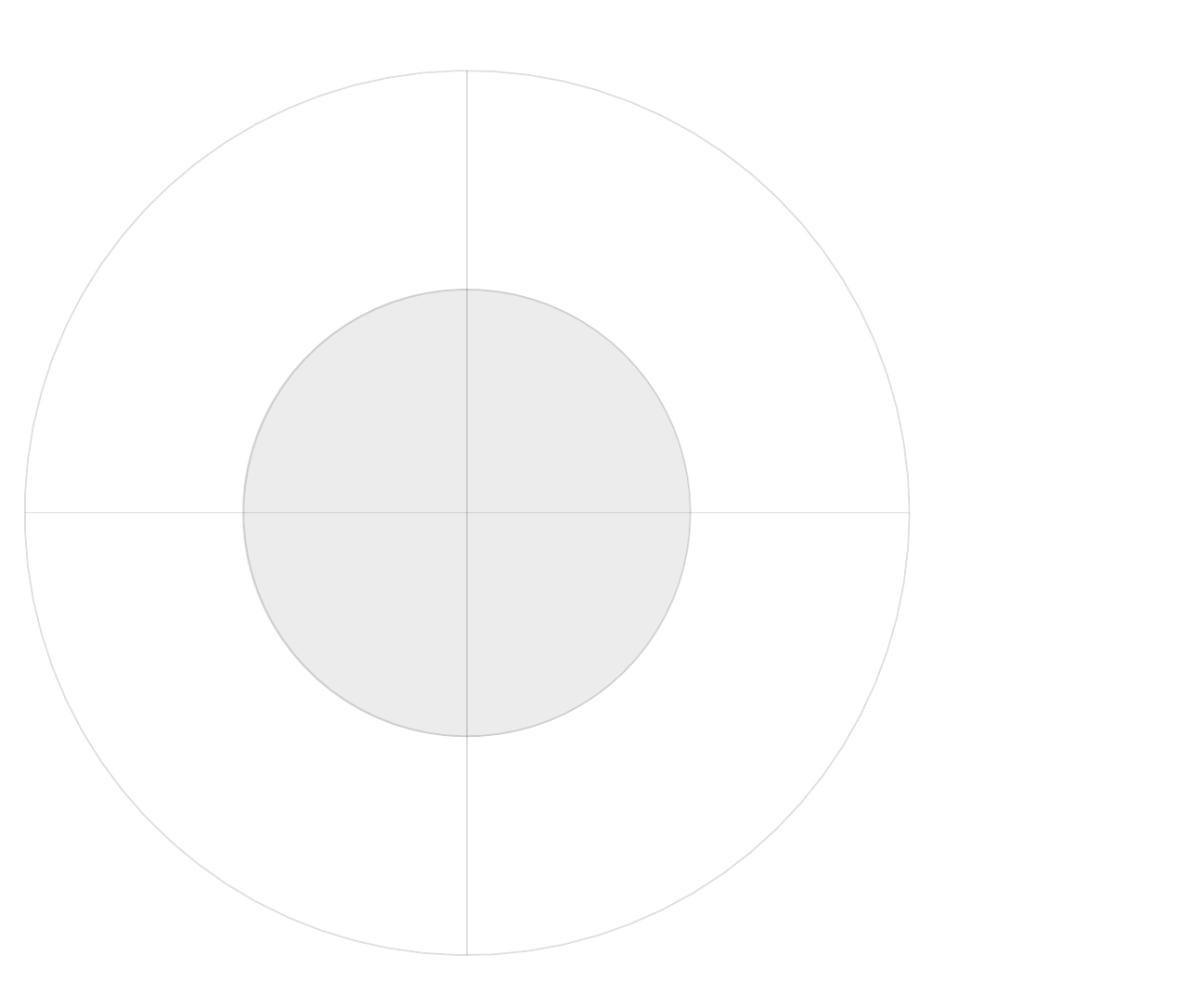
\caption{Polar plot showing the standard deviation for different quadrature directions in phase space. $\rho$ is a pure state squeezed along $q$ while $E$ is squeezed along $p$. The retrodiction is squeezed in both the $q$ and $p$ directions and is below the shot noise value of $\frac{1}{\sqrt{2}}$ for every quadrature.}
\label{fig:heisenberg}
\end{figure}

Squeezed Gaussian states of a single harmonic oscillator can be prepared by non-linear Hamiltonians or by monitoring of a specific quadrature by homodyne detection. If one quadrature is monitored until time $t$ and afterwards the conjugate quadrature is monitored, $\rho$ and $E$ may provide good estimates of both of these non-commuting observables of the system. The probability distribution of any linear combination of the two quadratures $x_\theta = q \cos\theta + p \sin\theta$ is a Gaussian and in the polar plot in Fig. \ref{fig:heisenberg} we show the standard deviation of this distribution as a function of the direction $\theta$ defining the quadrature observables. We observe that the retrodicted uncertainty is smaller in all directions than the predicted uncertainty by the density matrix $\rho$ alone, reflecting Eq. (\ref{eq:polarvariance}). We observe a Heisenberg ``butterfly'', reflecting the fact that the uncertainty may have minima along both the squeezed $q$ and the squeezed $p$ directions due to our probing of these observables before and after time $t$ respectively.

%\begin{figure}
%\centering
%\includesvg[width=\columnwidth,svgpath=images/,pretex=\footnotesize]{butterfly}
%\caption{Polar plot showing the standard deviation for different quadrature directions in phase space. $\rho$ is a pure state squeezed along $q$ while $E$ is squeezed along $p$. The retrodiction is squeezed in both the $q$ and $p$ directions and is below the shot noise value of $\frac{1}{\sqrt{2}}$ for every quadrature.}
%\label{fig:heisenberg}
%\end{figure}

\subsection{Unobserved evolution until a final projective measurement}

We know from the backward master equation for the effect matrix that if no measurement is made after $t$, $E$ is the identity matrix at all times, and our retrodiction does not differ from the predictions by the usual quantum mechanics formalism. If, however, a final measurement is carried out at time $T$, $E$ must be evolved backwards in time from the boundary condition set by the last measurement outcome.

In \cite{tan_homodyne_2017} the state of a superconducting qubit subject to a projective measurement at a final time $T$ was shown to differ appreciably from the usual exponential decay law, with consequences for retrodicted measurements on the system. Let us consider a similar example with a single oscillator mode which at $t=0$ occupies a coherent state $\ket{\alpha}$ and decays with rate $\Gamma$ into the ground state.

The master equation of the system is
\begin{equation}
\dd{\rho} = \mathcal{D}[\sqrt{\Gamma}a] \rho\dd{t}.
\end{equation}
The drift and the diffusion matrices can be calculated to be respectively $A = -\frac{\Gamma}{2} \idop$ and $ D =\Gamma \idop$ and thus the evolution of the mean values and covariance matrix for $\rho$ is described by
\begin{equation}
\dv{\ang{{\vb r}}}{t} = -\frac{\Gamma}{2} \ang{{\vb r}}, \qquad \dv{\sigma}{t} = \Gamma (\idop - \sigma).
\end{equation}
Since the initial state is the coherent state $\ket{\alpha} = \ket{\alpha_1 + i \alpha_2}$, it is easy to show that the state of the system is a coherent state decaying with mean values $\ang{{\vb r}}(t) = (\alpha_1,\alpha_2)\transp \sqrt{2} \exp{-\frac{\Gamma}{2}t}$ and constant covariance matrix, $\sigma(t)= \idop$.

Suppose that at time $T$ a projective measurement is performed that projects the oscillator into the (Gaussian) ground state. The mean values and covariance matrix equations for $E$ read 

\begin{equation} \label{eq:enon}
\frac{\mean{{\vb r}}(t-dt)- \mean{{\vb r}}(t)}{\dd{t}} = \frac{\Gamma}{2}\mean{{\vb r}}, \qquad \frac{\gamma(t-dt)-\gamma(t)}{\dd{t}} = \Gamma (\idop + \gamma).
\end{equation}
The final projection onto the ground state sets the boundary conditions for the effect matrix to $\mean{r}(T)=(0,0)\transp$ and $\gamma(T)=\idop$. The solution of Eq. (\ref{eq:enon}) yields \footnote{This result is in agreement with an analysis in the Fock state basis by Bruno Peaudecerf and Th\'eo Rybarczyk, \url{https://tel.archives-ouvertes.fr/tel-01078348/}}
\begin{equation}
\mean{{\vb r}}(t) = \begin{pmatrix}
0 \\ 0
\end{pmatrix},
\qquad
\gamma(t)=\idop\left(2e^{\Gamma(T-t)} -1 \right).
\end{equation}

If at some intermediate time $t$ the position quadrature was measured, we know that the past probability distribution is a Gaussian with the mean value $q_\text{p}$ and variance $\operatorname{Var}(q_\text{p})=\Delta(q_\text{p})/2$ given by Eq. (\ref{eq:pastprediction}):
\begin{equation}
\begin{split}
q_\text{p}(t) &= \sqrt{2}\alpha_1 \left( e^{-\frac{\Gamma}{2}t} -\frac{1}{2} e^{-\Gamma\left(T-\frac{t}{2}\right)} \right), \\
\Delta(q_\text{p},t) &= 1 - \frac{1}{2} e^{-\Gamma(T-t)}.
\end{split}
\end{equation}
Compared with the forward prediction of a simple decaying coherent state, the retrodicted average and variance are lower the closer we are to the post-selection time $T$.

It is clear with this example that even without continuously probing the system, but using only post-selection by a final measurement, the past quantum state can give a result significantly different from the conventional quantum state. This difference was first analyzed for a pre- and postselected system in \cite{aharonov_time_1964}, providing the so called Aharonov-Bergmann-Lebowitz rule for the probabilities of projective measurement results. The ABL rule was used in \cite{albert_curious_1985,aharonov_how_1988} to study counterintuitive statistics for intermediate measurements and to define weak values. The past quantum state formalism in simple cases reduces to the results of the ABL rule and to the weak value formalism when applied to the pertaining pre- and postselection scenarios \cite{tan_prediction_2015, tan_quantum_2016, tan_homodyne_2017}.

\section{Conclusion}
\label{sec:conclusion}

We have in this article presented a Gaussian state formalism that accounts for our ability to retrodict the outcome of measurements on a quantum system subject to dynamical evolution and probing before and after those measurements. We showed that the stochastic master equation for the density matrix and its adjoint version for an effect matrix $E$ can both be replaced by simpler equations for vectors of stochastically evolving first moments and matrices of deterministically evolving second moments of the Gaussian phase space functions.

The retrodicted evolution of the properties of a physical system may offer insight and allows precision estimation of external influences on both classical and quantum systems \cite{petersen_estimation_2006,gammelmark_hidden_2014,rybarczyk_forward-backward_2015,campagne-ibarcq_observing_2014,tsang_optimal_2010, tsang_fundamental_2012}. Given the many physical systems that are exactly or approximately described by Gaussian states, and the ability to incorporate a very large number of oscillator modes in the calculation at low cost, we believe that our theory may find wide applications.

While  the first and second moments of the $\rho$ and $E$ phase space distributions do exhaust our knowledge about the system, it is still a practical challenge to extract general information from our theory, such as the retrodicted expectation value of observables that are not first or second order in the quadrature coordinates. The Wigner and related phase space functions allow calculation of expectation values of suitably symmetrized observables by simple integration of complex functions over phase space. A practical recipe to evaluate the retrodicted mean values of suitably ordered functions of $q$ and $p$ for Gaussian states constitutes an ambitious and interesting goal for further studies.

\begin{acknowledgments}
The authors acknowledge financial support from the Villum Foundation.
\end{acknowledgments}

\appendix*

\section{Derivation of Gaussian evolution equation}
\label{sec:derivation}

\subsection{Forward evolution}

Let us consider a system of $n$ oscillators with canonical quadrature operators $\vb{r}\transp = (q_1,p_1,\dots,q_n,p_n)$ that undergoes an evolution described by master equation (\ref{eq:lindblad}), and let us assume a quadratic Hamiltonian $H = \frac{1}{2} \vb{r}\transp R \vb{r} = \sum_{jk} {r}_j R_{jk}{r}_k$ where $R$ is a real and symmetric matrix. The $m$ linear dissipation modes are grouped into a vector $\vb{c}\transp=(c_1,\dots,c_m)$ that can be written as $\vb{c} = \ct \vb{r}$, where $\ct$ is a $m\times n$ complex matrix.

Under such conditions an initially Gaussian state stays Gaussian. In order to study the evolution of the system we can then study only the first and second moments of the state.

We can calculate the evolution of the first moment vector by $\dd{\ang{r_i}} = \tr{r_i\dd{\rho}}$, where $\dd{\rho}$ is given by Eq. (\ref{eq:lindblad}). There are three main parts in the master equation, a Hamiltonian contribution, the dissipation terms and the measurement terms. We will go through the parts and summarize the calculations.

The Hamiltonian contribution is given by
\begin{equation}
\begin{split}
-i&\tr{\comm{H}{\rho}r_i} = -i\tr{\rho\comm{r_i}{H}} = \\
&= -\frac{i}{2} \sum_{jk} R_{jk} \ang{\comm{r_i}{r_j r_k}} = \\
&= -\frac{i}{2} \sum_{jk} R_{jk} \left( \ang{r_j \comm{r_i}{r_k}} + \ang{\comm{r_i}{r_j} r_k} \right) = \\
&= -\frac{i}{2} \sum_{jk} R_{jk} \tr{i\rho \left(r_j \Omega_{ik} + r_k \Omega_{ij}\right)} = \\
&= -\sum_{jk} \Omega_{ik} R_{kj} \ang{r_j},
\end{split}
\end{equation}
since $R$ is symmetric and thus $R_{jk} = R_{jk}$.

For the dissipation part let us consider first a single dissipation channel $h$. By looking at a single component of the vector of dissipation modes $\vb{c}$ we can write
\begin{equation}
\label{eq:cdefinition}
{c}_{h} = \sum_k \ct_{hk}{r}_{k},
\qquad
{c}_{h}^\dagger = \sum_l \ct_{hl}^* {r}_{l}.
\end{equation}
The dissipion part for mode $h$ can then be calculated to be
\begin{equation}
\begin{split}
&\Tr(\left( c_h\rho c_h^\dagger-\frac{1}{2} \acomm{c_h^\dagger c_h}{\rho}\right)r_i) = \\
&=\sum_{kl}\Tr(\ct_{hk}\ct_{hl}^* \left(r_l r_i r_k -\frac{1}{2}\acomm{r_i}{r_l r_k}\right) \rho) = \\
&=\frac{i}{2}\sum_{kl} \Omega_{il} \left( \ct_{hk}^* \ct_{hl} - \ct_{hk} \ct_{hl}^* \right) \ang{r_k},
\end{split}
\end{equation}
where we have used the antisymmetric property of the symplectic matrix $\Omega$. Summing over all the dissipation modes $h$ we get
\begin{equation}
\begin{split}
&\sum_{hkl} \Omega_{il}\frac{1}{2i} \left(  \ct_{hl}^* \ct_{hk} - \ct_{hl} \ct_{hk}^* \right) \ang{r_k}=\\
%&\sum_{hkl} \Omega_{il} \frac{1}{2i} \left(  [\ct^\dagger]_{lh} \ct_{hk} - [\ct\transp]_{lh}\ct_{hk}^* \right) \ang{r_k} \\
=&\sum_{kl} \Omega_{il} \frac{1}{2i} \left( [\ct^\dagger\ct]_{lk} - [\ct\transp \ct^*]_{lk}\right) \ang{r_k} = \\
=&\sum_{kl} \Omega_{il} \left[ \Im{\ct^\dagger\ct} \right]_{lk}\ang{r_k},
\end{split}
\end{equation}
since $\ct_{hl}^* = [\ct^\dagger]_{lh}$ and $\ct_{hl}=[\ct\transp]_{lh}$.

Finally for the measurement contribution let us define $a_{hk} = \Re{\ct_{hk}}$, $b_{hk} = \Im{\ct_{hk}}$. Considering a single measurement channel $h$:
\begin{equation}
\begin{split}
&\Tr( (c_h - \ang{c_h}) \rho r_i + \rho (c_h^\dagger -\ang{c_h^\dagger})r_i ) = \\
=&\sum_{k} a_{hk} \left( \ang{r_i r_k} + \ang{r_k r_i} -2\ang{r_i} \ang{r_k} \right) + i b_{hk}  \left( \ang{r_i r_k} - \ang{r_k r_i} \right)=  \\
=&\sum_{k} a_{hk} \sigma_{ik} - b_{hk}\Omega_{ik} = \sigma_{ik}\Re{\ct\transp}_{kh} - \Omega_{ik}\Im{\ct\transp}_{kh}.
\end{split}
\end{equation}

Combining all the contributions one can see that the evolution of the first moment is
\begin{equation}
\label{eq:rhofirst}
\dd{\ang{\vb{r}}} = A\ang{\vb{r}}\dd{t} + \left( \sigma B\transp -N\transp \right)\sqrt{\bm{\eta}}\dd{\vb{W}},
\end{equation}
where $A = \Omega(R + \Im{\ct^\dagger \ct})$, $B =\Re{\ct}$, and  $N = \Omega \Im{\ct}$.

The derivation of the evolution for the covariance matrix $\sigma$ follows the same structure. Some attention must be paid to the fact that, due to the stochastic term of the evolution, second order differentials must be taken into account in accordance with It\^o calculus, therefore for every component of the covariance matrix:
\begin{equation}
\label{eq:dsigma}
\begin{split}
\dd{\sigma_{ij}} = &\Tr(\dd{\rho}\left( r_i r_j + r_j r_i \right))+ \\
&-2\dd{\ang{r_i}}\ang{r_j} -2\ang{r_i}\dd{\ang{r_j}} -2 \dd{\ang{r_i}}\dd{\ang{r_j}}.\\
\end{split}
\end{equation}
Explicit calculations shows that all stochastic contributions cancel when we assume that the Wigner function is Gaussian and we apply Wick's theorem to third order moments as 
\begin{equation}
\ang{r_i r_j r_k} = \ang{r_i}\ang{r_j r_k} -\ang{r_i r_k}\ang{r_j} + \ang{r_i r_j}\ang{r_k}.
\end{equation}

Let us consider the first term in Eq. (\ref{eq:dsigma}) and study only $\Tr(\dd{\rho}r_i r_j)$ since it is symmetric in $i$ and $j$.

The Hamiltonian contribution is
\begin{equation}
\begin{split}
&-i\Tr(\comm{H}{\rho}r_i r_j) = -\frac{i}{2} \sum_{kl}R_{kl} \Tr(\rho\comm{r_i r_j}{r_k r_l}) = \\
=&\sum_{kl} R_{kl} \left( \ang{r_i r_k}\Omega_{jl} + \ang{r_k r_j}\Omega_{il} \right).
\end{split}
\end{equation}

The dissipation term reads
\begin{equation}
\begin{split}
&\Tr(\left( c_h\rho c_h^\dagger-\frac{1}{2} \acomm{c_h^\dagger c_h}{\rho}\right)r_i r_j) = \\
=&\frac{i}{2} \sum_{kl} \ct_{hk} \ct_{hl}^* ( \ang{r_i r_k}\Omega_{lj} -\ang{r_j r_k}\Omega{il} + \\
& \hphantom{\sum_{kl}\frac{i}{2} \ct_{hk} \ct_{hl}}-\ang{r_l r_i}\Omega_{kj} + \ang{r_l r_j}\Omega_{ik}).\\
\end{split}
\end{equation}

One has then to sum over all dissipation channels $h$ and the symmetric terms exchanging $i\leftrightarrow j$ to obtain the complete $\Tr(\dd{\rho}(r_i r_j + r_j r_i))$.

Considering the full variation given by Eq. (\ref{eq:dsigma}), the evolution for matrix $\sigma$ can be written in the following matricial form
\begin{equation}
\label{eq:ariccati}
\dv{\sigma}{t} = A\sigma +\sigma A\transp +D - 2\left( \gamma B\transp - N\transp\right)\bm{\eta}\left( \gamma B\transp - N \transp\right)\transp,
\end{equation}
where the diffusion matrix is defined as $D = -2\Omega\Re{\ct^\dagger \ct}\Omega$.
The last non linear term comes from the second order contribution $-2\dd{r_i}\dd{r_j}$ and the unmonitored evolution can be obtained by setting the efficiency matrix $\bm{\eta}=0$ and recover a linear evolution for $\sigma$.

\subsection{Backward evolution}

The derivation of the backward evolution of a Gaussian Wigner function of the effect matrix $E$ has many similarities with the derivation for the forward evolution of $\ang{{\vb r}}$ and $\sigma$.

The evolution of the first moment $\mean{r_i} \equiv \Tr(r_i E)$ is given by $\dd{r_i}=\Tr(r_i \dd{E})$ where $\dd{E}$ is given by the backward master equation (\ref{eq:blindblad}) which is to be compared with forward master equation (\ref{eq:lindblad}).

It is clear from the master equations that the Hamiltonian contribution differs only by a sign, therefore
\begin{equation}
i \tr{\comm{H}{E}r_i} = \sum_{jk}\Omega_{ik}R_{kj}\mean{r_j}.
\end{equation}

In the backward master equation the dissipation contribution is given by the adjoint dissipation superoperators. Making the master equation trace-preserving gives in the end a term that is the same in the forward evolution case and differs only by a sign:
\begin{equation}
\begin{split}
&\tr{\left(c_h^\dagger E c_h -\frac{1}{2}E c_h^\dagger c_h - (\mean{c_h c_h^\dagger} - \mean{c_h^\dagger c_h})E\right)r_i} = \\
&=-\frac{i}{2}\sum_{kl} \Omega_{il} \left( \ct_{hk}^* \ct_{hl} - \ct_{hk} \ct_{hl}^* \right) \mean{r_k}.
\end{split}
\end{equation}

Comparing the measurement terms between the forward and backward equation, since the difference is an exchange of $c_h \leftrightarrow c_h^\dagger$, the final result corresponds to changing the sign for the $b_{hk} = \Im{\ct_{hk}}$ terms, in fact
\begin{equation}
\begin{split}
&\tr{\left(c_h^\dagger - \mean{c_h^\dagger}\right)E r_i + E \left(c_h - \mean{c_h}\right)r_i } = \\
=& \sum_{k} a_{hk}\gamma_{ik} + b_{hk}\Omega_{ik} = \gamma_{ik}\Re{\ct\transp}_{kh} + \Omega_{ik}\Im{\ct\transp}_{kh}.
\end{split}
\end{equation}

For the evolution of the first moments of the effect matrix we have therefore
\begin{equation}
\label{eq:first}
\begin{split}
\dd{\mean{\vb{r}}} = &-A \mean{\vb{r}} \dd{t} + \left(\gamma B\transp + N\transp\right) \sqrt{\bm{\eta}} \dd{\vb{s}},
\end{split}
\end{equation}
where the drift matrix $A$, $B$, and $N$ are exactly the same matrices defined for the forward evolution of $\ang{\vb{r}}$ given in Eq. (\ref{eq:rhofirst}).

As regards the covariance matrix $\gamma$, second order terms must be again taken into account and $\dd{\gamma_{ij}}$ is calculated via an equation similar to Eq. (\ref{eq:dsigma}). The evolution of $\gamma$ is also deterministic in a similar way to $\sigma$, as the stochastic terms of the evolution cancel.

The Hamiltonian term differs again only by a sign, while the calculation of the effect of the adjoint dissipation superoperators gives
\begin{equation}
\begin{split}
&\tr{\left(c_h^\dagger E c_h -\frac{1}{2}E c_h^\dagger c_h - (\mean{c_h c_h^\dagger} - \mean{c_h^\dagger c_h})E\right)r_i r_j} = \\
=& \frac{i}{2} \sum_{kl} \ct_{hk} \ct_{hl}^* ( -\mean{r_k r_i}\Omega_{lj} + \mean{r_k r_j}\Omega_{il} + \\
&\hphantom{\frac{i}{2} \sum_{kl} \ct_{hk} \ct_{hl}^* -} +\mean{r_i r_l} \Omega_{kj} - \mean{r_j r_l}\Omega_{ik} ).
\end{split}
\end{equation}

Finally the evolution of $\gamma$ is given by
\begin{equation}
\label{eq:backwardriccati}
\begin{split}
\dv{\gamma}{t} = &-A\gamma - \gamma A\transp + D - 2\left( \gamma B\transp + N\transp\right)\bm{\eta}\left( \gamma B\transp + N \transp\right)\transp,
\end{split}
\end{equation}
where the diffusion matrix is the same defined for the forward Riccati equation given in Eq. (\ref{eq:ariccati}).

Some features of the $\gamma$ Riccati equation can be intuitively derived from the differences in the evolution of the first moments: $\gamma$ evolves with opposite drift matrix $A$ and the only change in the measurement contribution is a sign for the $N=\Omega\Im{\ct\transp}$ term. As in the forward in time case, it is the introduction of measurements that generates non-linear terms in the evolution of the covariance matrix and we can also see that measuring the system contributes by reducing the covariance matrix elements of $E$ backward in time.

%\bibliography{gaussianPQS}
\input{gaussianPQS.bbl}
\end{document}

%% file: system.pdf_tex
%% Creator: Inkscape inkscape 0.92.2, www.inkscape.org
%% Accompanies image file 'system.pdf' (pdf, eps, ps)
\begingroup%
  \makeatletter%
  \providecommand\color[2][]{%
    \errmessage{(Inkscape) Color is used for the text in Inkscape, but the package 'color.sty' is not loaded}%
    \renewcommand\color[2][]{}%
  }%
  \providecommand\transparent[1]{%
    \errmessage{(Inkscape) Transparency is used (non-zero) for the text in Inkscape, but the package 'transparent.sty' is not loaded}%
    \renewcommand\transparent[1]{}%
  }%
  \providecommand\rotatebox[2]{#2}%
  \ifx\svgwidth\undefined%
    \setlength{\unitlength}{342.12866115bp}%
    \ifx\svgscale\undefined%
      \relax%
    \else%
      \setlength{\unitlength}{\unitlength * \real{\svgscale}}%
    \fi%
  \else%
    \setlength{\unitlength}{\svgwidth}%
  \fi%
  \global\let\svgwidth\undefined%
  \global\let\svgscale\undefined%
  \makeatother%
  \begin{picture}(1,0.29664128)%
    \put(0,0){\includegraphics[width=\unitlength,page=1]{system.pdf}}%
    \put(0.34132247,0.2031177){\color[rgb]{0,0,0}\makebox(0,0)[lt]{\begin{minipage}{0.20343223\unitlength}\raggedright  \end{minipage}}}%
    \put(0.34413817,0.20398514){\color[rgb]{0,0,0}\makebox(0,0)[lt]{\begin{minipage}{0.19841146\unitlength}\raggedright  \end{minipage}}}%
    \put(0.33752444,0.18001049){\color[rgb]{0,0,0}\makebox(0,0)[lt]{\begin{minipage}{0.41170375\unitlength}\raggedright  \end{minipage}}}%
    \put(0.51524714,0.04985735){\color[rgb]{0,0,0}\rotatebox{90}{\makebox(0,0)[lt]{\begin{minipage}{0.17278341\unitlength}\raggedright Measurement\\ signal\end{minipage}}}}%
    \put(0.32302458,0.23996951){\color[rgb]{0,0,0}\makebox(0,0)[lt]{\begin{minipage}{0.21990609\unitlength}\raggedright Monitoring\end{minipage}}}%
    \put(0.01974798,0.0990755){\color[rgb]{0,0,0}\makebox(0,0)[lt]{\begin{minipage}{0.34060631\unitlength}\raggedright Open system of oscillators\end{minipage}}}%
    \put(0.78839846,0.02693566){\color[rgb]{0,0,0}\makebox(0,0)[lb]{\smash{$t$}}}%
    \put(0.88644799,0.02596703){\color[rgb]{0,0,0}\makebox(0,0)[lb]{\smash{$T$}}}%
    \put(0,0){\includegraphics[width=\unitlength,page=2]{system.pdf}}%
    \put(0.92125605,0.06632713){\color[rgb]{0,0,0}\makebox(0,0)[lb]{\smash{Time}}}%
    \put(0.60846182,0.04871426){\color[rgb]{0,0,0}\makebox(0,0)[lt]{\begin{minipage}{0\unitlength}\raggedright \end{minipage}}}%
    \put(0,0){\includegraphics[width=\unitlength,page=3]{system.pdf}}%
  \end{picture}%
\endgroup%

%% file: plan.tex
%\documentclass[tikz]{standalone}
%\usepackage{tikz}
%\usetikzlibrary{arrows,positioning}
%
%\begin{document}
\tikzset{state1/.style={draw, rounded corners, align=center, rectangle, inner xsep=1.5pt}}
\tikzset{state2/.style={draw, rounded corners, align=center, rectangle,inner xsep=1.5pt}}
\tikzset{state3/.style={draw, rounded corners, align=center, rectangle,inner xsep=1.5pt}}
\tikzset{evolution/.style={align=center, scale=0.85}}

\begin{tikzpicture} [node distance = 0.06\columnwidth, >=latex', every node/.style={minimum width=0cm}]

% Coordinates grid to help with positioning
%\draw[help lines,step=1,black!10] (0,0) grid (10,10);

\node (a) at (0,5) [state1] {$\rho$ \\[.3em]
\begin{tikzpicture}
\node [evolution] { Lindblad \\ master equation};
\end{tikzpicture}};

\node (b) [right=of a, state2] {$W_\rho({\bf r})$\\[.3em]
\begin{tikzpicture}
\node [evolution,rectangle] {Phase space\\evolution};
\end{tikzpicture}};

\node (c) [right=of b, state3] {$\ang{{\bf r}}, \sigma$ \\[.3em] 
\begin{tikzpicture}
%\node [evolution] {Stochastic first moment and second moment Riccati equation};
\node [evolution] {Stochastic first moment \\ and second moment \\ Riccati equation};
\end{tikzpicture}};

\node (d) [below=1cm of a, state1] {$E$\\[0.3em]
\begin{tikzpicture}
\node [evolution] {Backward \\ master equation};
\end{tikzpicture}};

\node (e) [right=of d, state2] {$W_E({\bf r})$ \\[0.3em]
\begin{tikzpicture}
\node [evolution] {Phase space \\ evolution};
\end{tikzpicture}};

\node (f) [right=of e, state3] {$\mean{{\bf r}}, \gamma$ \\[.3em] 
\begin{tikzpicture}
\node [evolution] {Stochastic first moment \\ and second moment \\ Riccati equation};
\end{tikzpicture}};

\foreach \x/\y in {a/b, b/c, a/d, d/e}
\draw [<->] (\x)--(\y);
\draw [<->] (e)--(f);
\end{tikzpicture}   

%\end{document}

%% file: phasespace.pdf_tex
%% Creator: Inkscape inkscape 0.92.2, www.inkscape.org
%% Accompanies image file 'phasespace.pdf' (pdf, eps, ps)
\begingroup%
  \makeatletter%
  \providecommand\color[2][]{%
    \errmessage{(Inkscape) Color is used for the text in Inkscape, but the package 'color.sty' is not loaded}%
    \renewcommand\color[2][]{}%
  }%
  \providecommand\transparent[1]{%
    \errmessage{(Inkscape) Transparency is used (non-zero) for the text in Inkscape, but the package 'transparent.sty' is not loaded}%
    \renewcommand\transparent[1]{}%
  }%
  \providecommand\rotatebox[2]{#2}%
  \ifx\svgwidth\undefined%
    \setlength{\unitlength}{482.95781654bp}%
    \ifx\svgscale\undefined%
      \relax%
    \else%
      \setlength{\unitlength}{\unitlength * \real{\svgscale}}%
    \fi%
  \else%
    \setlength{\unitlength}{\svgwidth}%
  \fi%
  \global\let\svgwidth\undefined%
  \global\let\svgscale\undefined%
  \makeatother%
  \begin{picture}(1,1.00667914)%
    \put(0,0){\includegraphics[width=\unitlength,page=1]{phasespace.pdf}}%
    \put(0.22177284,0.00419932){\makebox(0,0)[lb]{\smash{$\ang{q}$}}}%
    \put(0.48589714,0.00419932){\makebox(0,0)[lb]{\smash{$\mean{q}$}}}%
    \put(0,0){\includegraphics[width=\unitlength,page=2]{phasespace.pdf}}%
    \put(0.00040515,0.43591402){\makebox(0,0)[lb]{\smash{$\ang{p}$}}}%
    \put(0.0283579,0.23403307){\makebox(0,0)[lb]{\smash{$\mean{p}$}}}%
    \put(0,0){\includegraphics[width=\unitlength,page=3]{phasespace.pdf}}%
    \put(0.48986537,0.96288427){\color[rgb]{0,0,0}\makebox(0,0)[lb]{\smash{Density matrix marginal distr.}}}%
    \put(0.4903154,0.89870527){\color[rgb]{0,0,0}\makebox(0,0)[lb]{\smash{Past probability distribution}}}%
    \put(0,0){\includegraphics[width=\unitlength,page=4]{phasespace.pdf}}%
    \put(0.48943921,0.95044018){\color[rgb]{0,0,0}\makebox(0,0)[lt]{\begin{minipage}{0.54010357\unitlength}\raggedright Effect matrix marginal distr.\end{minipage}}}%
    \put(0.40726307,0.95141563){\color[rgb]{0,0,0}\makebox(0,0)[lb]{\smash{ }}}%
    \put(0,0){\includegraphics[width=\unitlength,page=5]{phasespace.pdf}}%
    \put(0.57907846,0.57386113){\color[rgb]{0,0,0}\makebox(0,0)[lb]{\smash{ }}}%
    \put(0.71717839,0.68090242){\color[rgb]{0,0,0}\makebox(0,0)[lb]{\smash{ }}}%
    \put(0.06273777,0.91826796){\color[rgb]{0,0,0}\makebox(0,0)[lt]{\begin{minipage}{0.21429401\unitlength}\raggedright Position \\ distribution\\  \end{minipage}}}%
    \put(0.42711313,1.00923632){\color[rgb]{0,0,0}\makebox(0,0)[lb]{\smash{ }}}%
    \put(0.70063498,0.66828758){\color[rgb]{0,0,0}\makebox(0,0)[lt]{\begin{minipage}{0.27148022\unitlength}\raggedright Momentum distribution\end{minipage}}}%
    \put(0.10519499,0.73181192){\color[rgb]{0,0,0}\makebox(0,0)[lb]{\smash{$\pr(q)$}}}%
    \put(0.27078635,0.87298077){\color[rgb]{0,0,0}\makebox(0,0)[lb]{\smash{$\pr_\text{p}(q)$}}}%
    \put(0.79786853,0.50590567){\color[rgb]{0,0,0}\makebox(0,0)[lb]{\smash{$\pr(p)$}}}%
    \put(0.87183376,0.33231866){\color[rgb]{0,0,0}\makebox(0,0)[lb]{\smash{$\pr_\text{p}(p)$}}}%
    \put(0.26451162,0.35374209){\color[rgb]{0,0,0}\makebox(0,0)[lb]{\smash{$W_\rho$}}}%
    \put(0.56099527,0.31923077){\color[rgb]{0,0,0}\makebox(0,0)[lb]{\smash{$W_E$}}}%
    \put(0.57348178,0.7208172){\color[rgb]{0,0,0}\makebox(0,0)[lb]{\smash{$\mu_E(q)$}}}%
    \put(0.79251263,0.15995226){\color[rgb]{0,0,0}\makebox(0,0)[lb]{\smash{$\mu_E(p)$}}}%
    \put(0.14626718,0.76232363){\color[rgb]{0,0,0}\makebox(0,0)[lb]{\smash{ }}}%
    \put(0.38631286,0.5719383){\color[rgb]{0,0,0}\makebox(0,0)[lt]{\begin{minipage}{0.28531181\unitlength}\raggedright  \end{minipage}}}%
    \put(0.38631286,0.5719383){\color[rgb]{0,0,0}\makebox(0,0)[lt]{\begin{minipage}{0.1662468\unitlength}\raggedright  \end{minipage}}}%
    \put(0.33297718,0.55330309){\color[rgb]{0,0,0}\makebox(0,0)[lt]{\begin{minipage}{0.33864755\unitlength}\centering Phase space\\ distribution\\  \end{minipage}}}%
    \put(0.47716689,0.3860465){\color[rgb]{0,0,0}\makebox(0,0)[b]{\smash{ }}}%
  \end{picture}%
\endgroup%

%% file: oscillator.pdf_tex
%% Creator: Inkscape inkscape 0.92.2, www.inkscape.org
%% Accompanies image file 'oscillator.pdf' (pdf, eps, ps)
\begingroup%
  \makeatletter%
  \providecommand\color[2][]{%
    \errmessage{(Inkscape) Color is used for the text in Inkscape, but the package 'color.sty' is not loaded}%
    \renewcommand\color[2][]{}%
  }%
  \providecommand\transparent[1]{%
    \errmessage{(Inkscape) Transparency is used (non-zero) for the text in Inkscape, but the package 'transparent.sty' is not loaded}%
    \renewcommand\transparent[1]{}%
  }%
  \providecommand\rotatebox[2]{#2}%
  \ifx\svgwidth\undefined%
    \setlength{\unitlength}{412.5bp}%
    \ifx\svgscale\undefined%
      \relax%
    \else%
      \setlength{\unitlength}{\unitlength * \real{\svgscale}}%
    \fi%
  \else%
    \setlength{\unitlength}{\svgwidth}%
  \fi%
  \global\let\svgwidth\undefined%
  \global\let\svgscale\undefined%
  \makeatother%
  \begin{picture}(1,1.18181818)%
    \put(0,0){\includegraphics[width=\unitlength,page=1]{oscillator.pdf}}%
    \put(0.12272727,0.81757582){\makebox(0,0)[lb]{\smash{0}}}%
    \put(0.24090909,0.81757582){\makebox(0,0)[lb]{\smash{0.5}}}%
    \put(0.38090909,0.81757582){\makebox(0,0)[lb]{\smash{1}}}%
    \put(0.49909091,0.81757582){\makebox(0,0)[lb]{\smash{1.5}}}%
    \put(0.63909091,0.81757582){\makebox(0,0)[lb]{\smash{2}}}%
    \put(0.75727273,0.81757582){\makebox(0,0)[lb]{\smash{2.5}}}%
    \put(0.89727273,0.81757582){\makebox(0,0)[lb]{\smash{3}}}%
    \put(0.5,0.78181818){\makebox(0,0)[lb]{\smash{$\Gamma t$}}}%
    \put(0,0){\includegraphics[width=\unitlength,page=2]{oscillator.pdf}}%
    \put(0.08121218,0.84272727){\makebox(0,0)[lb]{\smash{-10}}}%
    \put(0.09757582,0.90318182){\makebox(0,0)[lb]{\smash{-5}}}%
    \put(0.10484855,0.96363636){\makebox(0,0)[lb]{\smash{0}}}%
    \put(0.10484855,1.02409091){\makebox(0,0)[lb]{\smash{5}}}%
    \put(0.09030309,1.08454545){\makebox(0,0)[lb]{\smash{10}}}%
    \put(0.06727273,0.94909091){\rotatebox{90}{\makebox(0,0)[lb]{\smash{$\ang{q}$}}}}%
    \put(0,0){\includegraphics[width=\unitlength,page=3]{oscillator.pdf}}%
    \put(0.12272727,0.46303036){\makebox(0,0)[lb]{\smash{0}}}%
    \put(0.24090909,0.46303036){\makebox(0,0)[lb]{\smash{0.5}}}%
    \put(0.38090909,0.46303036){\makebox(0,0)[lb]{\smash{1}}}%
    \put(0.49909091,0.46303036){\makebox(0,0)[lb]{\smash{1.5}}}%
    \put(0.63909091,0.46303036){\makebox(0,0)[lb]{\smash{2}}}%
    \put(0.75727273,0.46303036){\makebox(0,0)[lb]{\smash{2.5}}}%
    \put(0.89727273,0.46303036){\makebox(0,0)[lb]{\smash{3}}}%
    \put(0.5,0.42727273){\makebox(0,0)[lb]{\smash{$\Gamma t$}}}%
    \put(0,0){\includegraphics[width=\unitlength,page=4]{oscillator.pdf}}%
    \put(0.08121218,0.48818182){\makebox(0,0)[lb]{\smash{-10}}}%
    \put(0.09757582,0.54863636){\makebox(0,0)[lb]{\smash{-5}}}%
    \put(0.10484855,0.60909091){\makebox(0,0)[lb]{\smash{0}}}%
    \put(0.10484855,0.66954545){\makebox(0,0)[lb]{\smash{5}}}%
    \put(0.09030309,0.73){\makebox(0,0)[lb]{\smash{10}}}%
    \put(0.06727273,0.60363636){\rotatebox{90}{\makebox(0,0)[lb]{\smash{$\mean{q}$}}}}%
    \put(0,0){\includegraphics[width=\unitlength,page=5]{oscillator.pdf}}%
    \put(0.12272727,0.10848491){\makebox(0,0)[lb]{\smash{0}}}%
    \put(0.24090909,0.10848491){\makebox(0,0)[lb]{\smash{0.5}}}%
    \put(0.38090909,0.10848491){\makebox(0,0)[lb]{\smash{1}}}%
    \put(0.49909091,0.10848491){\makebox(0,0)[lb]{\smash{1.5}}}%
    \put(0.63909091,0.10848491){\makebox(0,0)[lb]{\smash{2}}}%
    \put(0.75727273,0.10848491){\makebox(0,0)[lb]{\smash{2.5}}}%
    \put(0.89727273,0.10848491){\makebox(0,0)[lb]{\smash{3}}}%
    \put(0.5,0.07272727){\makebox(0,0)[lb]{\smash{$\Gamma t$}}}%
    \put(0,0){\includegraphics[width=\unitlength,page=6]{oscillator.pdf}}%
    \put(0.08121218,0.13363636){\makebox(0,0)[lb]{\smash{-10}}}%
    \put(0.09757582,0.19454545){\makebox(0,0)[lb]{\smash{-5}}}%
    \put(0.10484855,0.25545455){\makebox(0,0)[lb]{\smash{0}}}%
    \put(0.10484855,0.31636364){\makebox(0,0)[lb]{\smash{5}}}%
    \put(0.09030309,0.37727273){\makebox(0,0)[lb]{\smash{10}}}%
    \put(0.06727273,0.25272727){\rotatebox{90}{\makebox(0,0)[lb]{\smash{$q_\text{p}$}}}}%
    \put(0,0){\includegraphics[width=\unitlength,page=7]{oscillator.pdf}}%
  \end{picture}%
\endgroup%

%% file: butterfly.pdf_tex
%% Creator: Inkscape inkscape 0.92.2, www.inkscape.org
%% Accompanies image file 'butterfly.pdf' (pdf, eps, ps)
\begingroup%
  \makeatletter%
  \providecommand\color[2][]{%
    \errmessage{(Inkscape) Color is used for the text in Inkscape, but the package 'color.sty' is not loaded}%
    \renewcommand\color[2][]{}%
  }%
  \providecommand\transparent[1]{%
    \errmessage{(Inkscape) Transparency is used (non-zero) for the text in Inkscape, but the package 'transparent.sty' is not loaded}%
    \renewcommand\transparent[1]{}%
  }%
  \providecommand\rotatebox[2]{#2}%
  \ifx\svgwidth\undefined%
    \setlength{\unitlength}{373.26846313bp}%
    \ifx\svgscale\undefined%
      \relax%
    \else%
      \setlength{\unitlength}{\unitlength * \real{\svgscale}}%
    \fi%
  \else%
    \setlength{\unitlength}{\svgwidth}%
  \fi%
  \global\let\svgwidth\undefined%
  \global\let\svgscale\undefined%
  \makeatother%
  \begin{picture}(1,0.81477695)%
    \put(-0.99864701,0.77806942){\color[rgb]{0,0,0}\makebox(0,0)[lb]{\smash{}}}%
    \put(0.76658365,0.37934587){\makebox(0,0)[lb]{\smash{$q$}}}%
    \put(0,0){\includegraphics[width=\unitlength,page=1]{butterfly.pdf}}%
    \put(0.36305368,0.77483863){\makebox(0,0)[lb]{\smash{$p$}}}%
    \put(0.39017892,0.56118551){\makebox(0,0)[lb]{\smash{$\frac{1}{\sqrt{2}}$}}}%
    \put(0,0){\includegraphics[width=\unitlength,page=2]{butterfly.pdf}}%
    \put(0.68559188,0.48320136){\color[rgb]{0,0,0}\makebox(0,0)[lb]{\smash{$\theta$}}}%
    \put(0.61241945,0.77395133){\color[rgb]{0,0,0}\makebox(0,0)[lt]{\begin{minipage}{0.31561045\unitlength}\raggedright Standard deviation for quadrature $x_\theta$\end{minipage}}}%
    \put(0,0){\includegraphics[width=\unitlength,page=3]{butterfly.pdf}}%
    \put(0.7699744,0.07280517){\color[rgb]{0,0,0}\makebox(0,0)[lt]{\begin{minipage}{0.26413748\unitlength}\raggedright PQS retrodiction\end{minipage}}}%
    \put(0.77036848,0.13391237){\color[rgb]{0,0,0}\makebox(0,0)[lb]{\smash{$\rho$ }}}%
    \put(0.77036848,0.09192124){\color[rgb]{0,0,0}\makebox(0,0)[lb]{\smash{$E$}}}%
  \end{picture}%
\endgroup%